\documentclass[lettersize,journal]{IEEEtran}
\usepackage{amsmath,amsfonts}
\usepackage{amssymb}
\usepackage{algorithmic}
\usepackage{algorithm}
\usepackage{array}
\usepackage[caption=false,font=normalsize,labelfont=sf,textfont=sf]{subfig}
\usepackage{textcomp}
\usepackage{stfloats}
\usepackage{url}
\usepackage{verbatim}
\usepackage{graphicx}
\usepackage{multirow}
\usepackage{cite}
\usepackage{bm}
\usepackage{xcolor}
\begin{document}

\title{Cross-IP Request Coalescing: Relocating the Fan-out Point in Virtualized I/O}

\author{Kiseok Kim, Hyeontae Joo, and Hwangnam Kim, \IEEEmembership{Member, IEEE}

\thanks{Manuscript received November 00, 2000; revised February 00, 2000.}
\thanks{(Corresponding author: Hwangnam Kim) \\
 Kiseok Kim and Hyeontae Joo contributed equally to this work. \\
 Kiseok Kim, Hyeontae Joo, and Hwangnam Kim are with the Korea University, Seoul 02841, Republic of Korea (e-mail: \{kisuk528, motern800, and hnkim\}@korea.ac.kr)}
}

\markboth{IEEE Journal of \LaTeX\ Class,~Vol.~12, No.~6, February~2024}%
{Shell \MakeLowercase{\textit{et al.}}: A Sample Article Using IEEEtran.cls for IEEE Journals}

\IEEEpubid{0000-0000~\copyright~2024 IEEE}

\maketitle
\begin{abstract}
Cloud data centers rely on virtualization technologies to serve AI workloads in multi-tenant environments. With the growing scale of data-intensive AI workloads, the performance of storage I/O paths at the virtualization layer has become a critical factor. A single user request often crosses multiple IP blocks; functional units such as storage, GPU, and accelerator devices, under virtualization, fanning out into separate stack traversals between the guest and the backend. As a result, round-trip and context-switching overheads accumulate with the number of devices. In this letter, we identify that a dominant factor in this overhead lies not in the kernel-mediated I/O path alone, but in the per-device submission structure itself, which persists even in user-space, kernel-bypass storage frameworks such as SPDK. To address this, we propose cross-IP request coalescing, which relocates the fan-out point from the guest to the backend.
As a first instantiation, we apply this principle to block storage I/O. The guest submits a single coalesced request, and an SPDK acceleration block device (bdev) decomposes it and dispatches per-device I/O locally, so that $N$ device requests share a single stack traversal instead of $N$. Evaluation in an SPDK-based virtualized environment shows that the proposed approach achieves up to 1.78× lower latency than the per-device baseline, with the benefit growing as concurrency increases.
\end{abstract}


\section{Introduction}
\IEEEPARstart{V}{irtualization}
technologies, ranging from hypervisor-based virtual machines to container-based platforms, support cloud and AI infrastructure, yet their I/O paths incur overhead across multiple layers as workloads demand concurrent access to high-speed storage devices~\cite{yang2026here}.
    The rapid growth of data-intensive AI workloads also leads to concurrent requests across multiple devices~\cite{jayanetti2024multi}.
    As GPUs, memory, and storage scale up, the path overhead of data transfer has emerged as a critical issue in fully utilizing hardware performance.
    Among these issues, device I/O performance is one of the performance-critical paths, yet a residual inefficiency persists even in existing solutions~\cite{saber2025physical}.
    
    Existing techniques such as user-space I/O acceleration, hardware-assisted offloading, and device assignment reduce software overhead along the I/O path.
    However, they do not directly address the per-device submission overhead that scales with the number of devices a single request spans, as in replicated storage writes or multi-volume parallel writes, where a guest issues concurrent I/O across multiple devices~\cite{russell2008virtio}.
    
\begin{figure}[t]
\centering
\includegraphics[width=2.1in,height=1.0in]{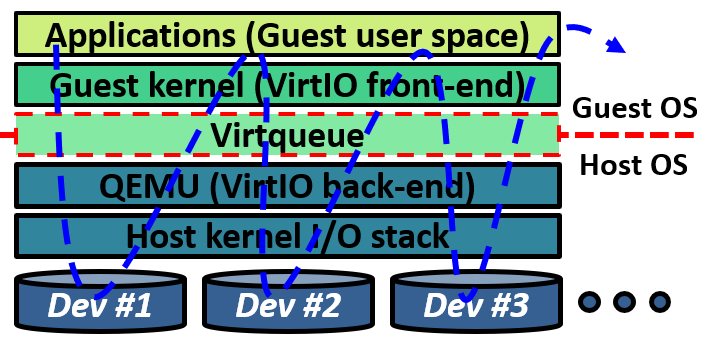}
\caption{Multi-device I/O fanning out across the virtualization stack.}
\label{fig-gen}
\end{figure}

    In this context, we analyze the structural overhead of the virtualized multi-device I/O path on QEMU, a representative type-2 hypervisor, and observe that repeated per-device traversals occur between the guest and the backend under concurrent multi-device access, as illustrated in Fig.~\ref{fig-gen}. We further observe that this overhead persists even with an SPDK vhost-user backend. Although user-space I/O stacks such as SPDK bypass the kernel path and provide benefits in high-speed storage I/O~\cite{torp2025path}, they leave the per-request submission structure intact, so the repetitive guest-to-backend overhead persists regardless of kernel bypass. Inspired by memory coalescing, which merges accesses that are separated only by address into a single transaction, we coalesce I/O requests separated by the virtualization structure across devices. We propose cross-IP request coalescing, which relocates the fan-out point of multi-device I/O from the guest into the SPDK vhost-user backend.

By relocating the fan-out point of multi-device I/O into the SPDK vhost-user backend, the accelerator 1)~eliminates the repeated guest-to-backend traversals, and 2)~confines context-switch accumulation and virtqueue contention to one submission instead of $N$.
To this end, we implement a custom bdev module in the SPDK backend and define a coalesced request format that carries multiple per-device I/Os under a single submission.

This letter makes the following contributions:
\begin{itemize}
    \item We identify a per-device submission bottleneck in the virtualized storage path of data-intensive multi-device workloads, and relocate the fan-out point from the guest into the SPDK vhost-user backend.
    \item We design a Virtio-based vhost-user device backed by an SPDK bdev, together with a coalesced
    request format that aggregates per-device I/O under a single batch header, exposed to the guest through the existing Virtio path without driver modification.
\end{itemize}

\IEEEpubidadjcol

\section{Background and Motivation}
\subsection{Virtualization I/O path in QEMU}

QEMU is a widely used type-2 hypervisor that executes as a user-space process on the host OS. As illustrated in Fig.~\ref{fig-gen}, a storage I/O request from a guest application flows through four functional layers: 1) the guest user space, 2) the guest kernel space including the VirtIO block driver, 3) the QEMU process comprising the VirtIO backend and device emulation, and 4) the host kernel I/O stack which reaches the physical storage devices.
Within the guest kernel, the VirtIO front-end driver enqueues I/O requests into shared memory (Virtqueue); VirtIO serves as a common abstraction layer over heterogeneous backend devices~\cite{park2022ambient}. The QEMU backend polls the Virtqueue, performs device emulation, and issues the corresponding I/O operations to the host OS's virtual filesystem (VFS) and block layers, ultimately reaching the physical devices.
This end-to-end path constitutes the north-south I/O stack that every storage request must traverse. When a guest issues concurrent requests, each request traverses this path separately, leading to redundant context switches and ring synchronization overhead.
\begin{table}[t]
\centering
\caption{Baseline I/O Latency by Number of Block Devices}
\label{tab:motive}
\renewcommand{\arraystretch}{0.9}
\begin{tabular}{ccc}
\hline
\textbf{Devices} & \textbf{Normal} [$\mu$s] & \textbf{High-load} [$\mu$s] \\
\hline
1 & 1.5  & 2.5  \\
2 & 3.5  & 6.0  \\
4 & 5.5  & 11.8 \\
8 & 7.8  & 24.4 \\
\hline
\end{tabular}
\end{table}
\subsection{Overhead characterization of multi-device I/O}
We characterize the I/O round-trip overhead under multi-device access, measured as the round-trip from guest issue to completion as a guest forwards requests across multiple devices. Each request is submitted separately and incurs guest-to-host context switches and virtqueue polling, so the latency accumulates with the number of concurrent submissions; accessing $N$ devices is a particular case of this accumulation, issuing $N$ separate submissions. As a baseline, we measure the round-trip I/O latency under two workload conditions, normal and high-load, while scaling the number of block devices from 1 to 8 as shown in Table~\ref{tab:motive}.
In the normal case, the overhead increases linearly with the number of devices. Under high load, however, the latency grows non-linearly, reaching 24.4~$\mu$s at eight devices, which we attribute to the cumulative impact of context switching and virtqueue contention. These results indicate that repeated per-request traversals, not the device operation itself, constitute a major bottleneck in multi-device virtualized I/O.

Crucially, the overhead arises not from the kernel path but from a submission structure handled separately per request. Even when the kernel path is bypassed, the guest-side submission remains per-request, so the same overhead persists~\cite{yang2017spdk}. Vhost-user offloads backend processing to a separate user-space process but is likewise bound by the per-request guest-side submission. Because the overhead is structural rather than path-specific, kernel bypass alone cannot remove it; the fan-out point itself must be relocated. Motivated by this, we propose a cross-IP request coalescing mechanism at the vhost-user backend (SPDK bdev) that coalesces the separate per-request submissions into a single coalesced submission, leaving the conventional block stack unchanged. The architectural details follow in the next section.

\begin{figure}[t]
\centering
\includegraphics[width=3.4in,height=1.5in]{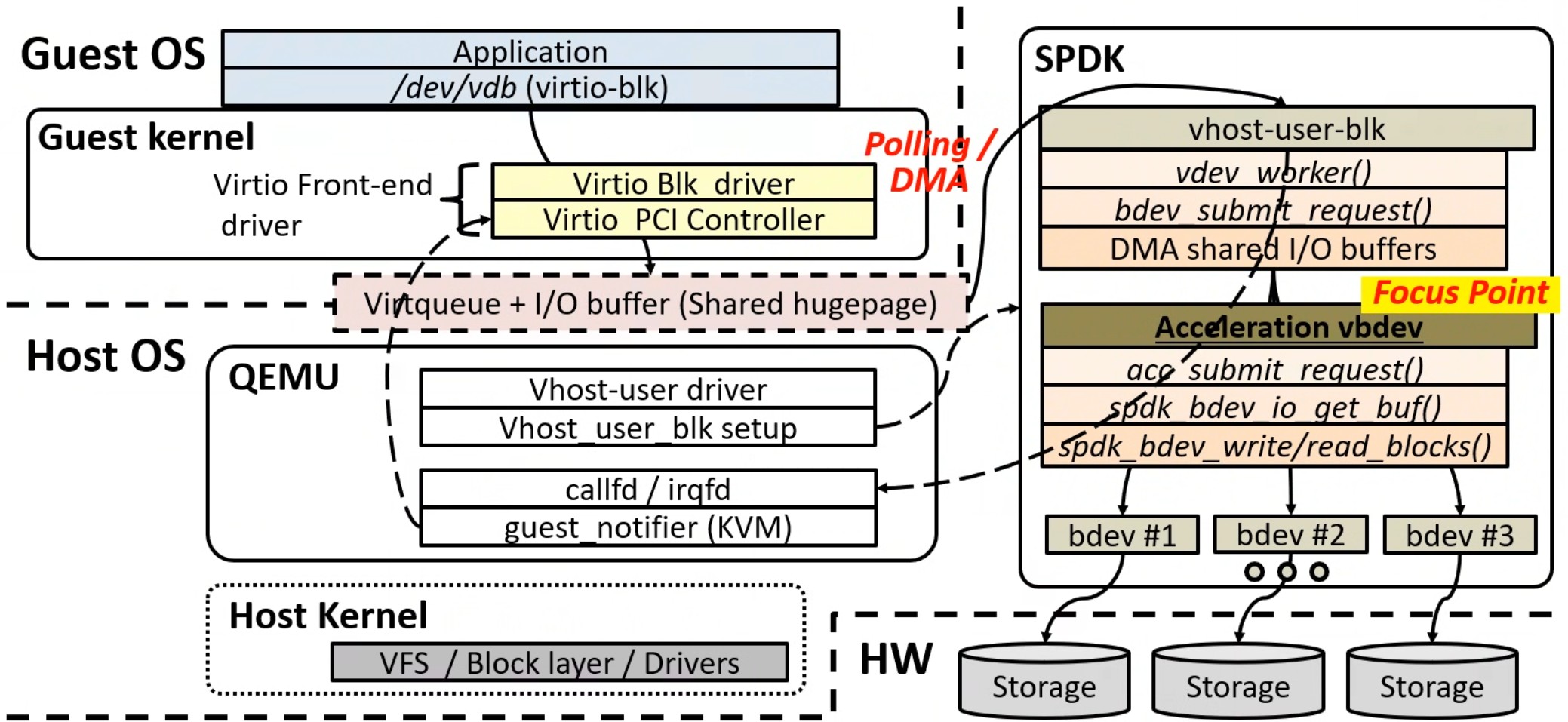}
\caption{Virtualized I/O hierarchy with proposed acceleration path.}
\label{fig-hier}
\end{figure}

\section{Design}
\subsection{Architecture Overview}
In multi-device I/O, the guest-side submission path is traversed once per request, repeated across the $N$ concurrent I/O requests issued to multiple block devices. This repeated path is dominated not by the device operation itself, but by the submission and control cost of delivering each request from the guest to the vhost-user backend. Each request separately passes through the guest-kernel block layer, the VirtIO front-end, the virtqueue, and the SPDK vhost-user backend, so the submission overhead accumulates as $N$ increases.

To address this, we do not modify the underlying block backend. Instead, we relocate the fan-out point of multi-device I/O from the guest side to the vhost-user backend in SPDK: the guest submits a single coalesced request carrying payloads for multiple devices, and the acceleration bdev decomposes it into per-device payload and dispatches each to the corresponding backend. From a control/data-plane view, the conventional path couples one control submission to each device's data path, whereas the relocated fan-out allows a single control submission to derive all $N$ transfers within the backend.

\subsection{Compound Request and Dispatch}
The compound request consists of a batch header followed by a sequence of per-device records. The header carries the number of records and the metadata needed to delimit each one; per record, the metadata holds the minimal information required by the SPDK bdev: a target device index, the payload size, and the payload data. The device index identifies the backend to which the payload is redirected, while the size and data fields define the I/O content to be delivered.

On the guest side, an application or runtime packs the payloads into a contiguous compound buffer, each tagged with its target device index, and submits the buffer once to the dedicated VirtIO-based acceleration device. The guest-visible submission thus remains a single request even though it internally represents multiple device-level I/O operations.

On the backend side, the acceleration bdev parses the request at the SPDK backend. It reads the batch header for the record count and layout, then iterates over the records, extracts each target device index and payload boundary, constructs an internal I/O vector, and forwards the payload to the corresponding bdev. The device index resolves the existing backend device, letting the accelerator reuse the conventional bdev structure without modifying the backend device layer, while dispatching multiple per-device I/O requests concurrently from a single coalesced submission.

\subsection{Modeling and Implementation}
We decompose the virtualized I/O path into its constituent stages: guest-side submission, virtqueue transport, vhost-user backend entry, and backend block service. Then we analyze how each stage scales under fan-out of $N$ concurrent I/O requests. We model the submission direction only. The completion path is excluded, since the proposed path also completes once rather than $N$ times.  Each of the $N$ requests requires a guest-side submission $T_{\mathrm{sub}}$, virtqueue handling $T_{\mathrm{vq}}$, and vhost-user backend entry $T_{\mathrm{vhost}}$, with backend block service time $T_{\mathrm{blk},i}$ for the $i$-th request. Since both paths submit their backend I/O asynchronously and process them as in-flight requests, we denote the resulting backend service time by $T_{\mathrm{blk}}(N) \approx \max_{1 \le i \le N} T_{\mathrm{blk},i}$, which is common to both paths. The baseline latency for multi-device I/O can then be bounded below as:
\begin{equation}
\label{eq-base}
T_{\mathrm{base}}(N) \ge N(T_{\mathrm{sub}} + T_{\mathrm{vq}} + T_{\mathrm{vhost}}) + T_{\mathrm{blk}}(N),
\end{equation}
where the bound is reached without contention, since lock and virtqueue contention raise the actual latency.  $N(T_{\mathrm{sub}} + T_{\mathrm{vq}} + T_{\mathrm{vhost}}) $ represents the guest-to-backend submission path.
Since the guest issues $N$ requests through the same submission context, this control path is repeatedly traversed, while $T_{\mathrm{blk}}(N)$ denotes the backend block service time.

In the proposed coalesced-dispatch path, the guest issues a single coalesced request rather than $N$ separate ones, so the per-request submission latency is incurred only once, in exchange for additional parsing and dispatch at the backend. $T_{\mathrm{parse}}(N)$ is the parsing latency of the coalesced request, and $T_{\mathrm{disp}}(N)$ the fan-out dispatch latency at the SPDK backend.
The proposed path relocates the fan-out into the backend: the same $N$-way distribution that the conventional path performs on the guest side, traversing the full north-south path $N$ times, is instead carried out locally within the vhost-user backend. Because the backend I/O is issued only after the coalesced request is parsed and dispatched, it enters the backend later than under per-request submission by an offset $\delta(N)$; we define the effective dispatch term $\tilde{T}_{\mathrm{disp}}(N) \triangleq T_{\mathrm{disp}}(N) + \delta(N)$, so that the backend service time $T_{\mathrm{blk}}(N)$ remains common to both paths. The latency of the proposed path can then be represented as:
\begin{equation}
\label{eq-prop}
\begin{split}
T_{\mathrm{comp}}(N) &= (T_{\mathrm{sub}} + T_{\mathrm{vq}} + T_{\mathrm{vhost}}) + T_{\mathrm{parse}}(N)\\
&\quad + \tilde{T}_{\mathrm{disp}}(N) + T_{\mathrm{blk}}(N).
\end{split}
\end{equation}

The submission and control latency $(T_{\mathrm{sub}} + T_{\mathrm{vq}} + T_{\mathrm{vhost}})$
occurs only once for the single coalesced submission, whereas the baseline repeats
it $N$ times. Subtracting $T_{\mathrm{comp}}(N)$ from the lower bound $T_{\mathrm{base}}(N)$
therefore eliminates the shared term $T_{\mathrm{blk}}(N)$. 
The proposed mechanism reduces the total latency whenever the parsing and dispatch overhead
is smaller than the repeated submission and control overhead it removes:
\begin{equation}
\label{eq-fit}
T_{\mathrm{parse}}(N) + \tilde{T}_{\mathrm{disp}}(N) < (N-1)\left( T_{\mathrm{sub}} + T_{\mathrm{vq}} + T_{\mathrm{vhost}} \right),
\end{equation}
where $\tilde{T}_{\mathrm{disp}}(N) \ge T_{\mathrm{disp}}(N)$ since it includes the offset $\delta(N)$.

Since $T_{\mathrm{base}}(N)$ is a lower bound, this condition is not necessary. Therefore, (\ref{eq-fit}) guarantees a benefit even without contention, and any lock or virtqueue contention in the baseline only enlarges the performance gap. 
The $(N-1)$ factor makes the latency reduction increasingly dominant as $N$ grows, since the conventional path repeats the full north-south path per request while the proposed path incurs parsing and dispatch only once, locally at the SPDK backend. If requests arrive stochastically, $N$ becomes a random variable determined by the arrival process, and both sides of~\eqref{eq-fit} scale approximately linearly with $N$. The expectation of the difference can be described as follows:
\begin{equation}
\label{eq-exp}
\begin{split}
E[T_{\mathrm{parse}}(N)] &+ E[\tilde{T}_{\mathrm{disp}}(N)] \\&< (E[N]-1)\left( T_{\mathrm{sub}} + T_{\mathrm{vq}} + T_{\mathrm{vhost}} \right),
\end{split}
\end{equation}
where the right-hand side follows from linearity of expectation. Given the same per-device I/O arrivals, the two paths differ in how the submissions combine. The baseline accumulates $N$ of them serially, so its submission latency follows a hypo-exponential distribution with mean $N/\mu$ for per-device submission rate $\mu$, whereas the proposed path incurs the submission cost only once, with mean $1/\mu$. The expected gap thus scales with $(E[N]-1)/\mu$, so the expected latency decreases with the mean number of coalesced requests.

We implement this mechanism as an acceleration bdev in the SPDK vhost-user backend, exposed to the guest as a VirtIO-based device. It issues the per-device backend I/O asynchronously and joins their completions before completing the coalesced request, so that the backend service time is determined by the slowest concurrent device. The design supports coordinated multi-device reads and writes. The design is not limited to QEMU, SPDK, or block storage but it applies wherever a single request fans out across multiple device-level operations (e.g., across storage, GPU, or accelerator devices), which we demonstrate on block storage as a first instantiation.

\section{Evaluation}
We evaluate the proposed fan-out relocation mechanism against a per-device baseline on a QEMU guest backed by an SPDK vhost-user target. The accelerator bdev and the baseline devices are configured with identical backend capacity, so that both paths transfer the same aggregate payload and differ only in the number of guest-to-backend round-trips.

In this setup, we consider two parameters: 1) the workload intensity (the number of concurrently in-flight requests coalesced into one submission, the experimental realization of $N$ in our model), and 2) the I/O size (the per-I/O payload size that dominates per-bdev processing at the SPDK backend). We vary the workload intensity ($1$-$32$ in-flight requests coalesced submission) and the I/O size ($1024$-$8192$~B), under both random-read and random-write workloads, with the backend device count fixed, so that the intensity maps directly to $N$, allowing us to test the $(N-1)$ scaling predicted by Eq.~\ref{eq-fit}. The guest uses a single virtqueue, so that intensity reflects in-flight concurrency rather than hardware queue parallelism. Each configuration averages $20$ runs of $10^3$ coalesced submissions.
\begin{figure}[t]
\centering
\includegraphics[width=3.8in, height=1.5in]{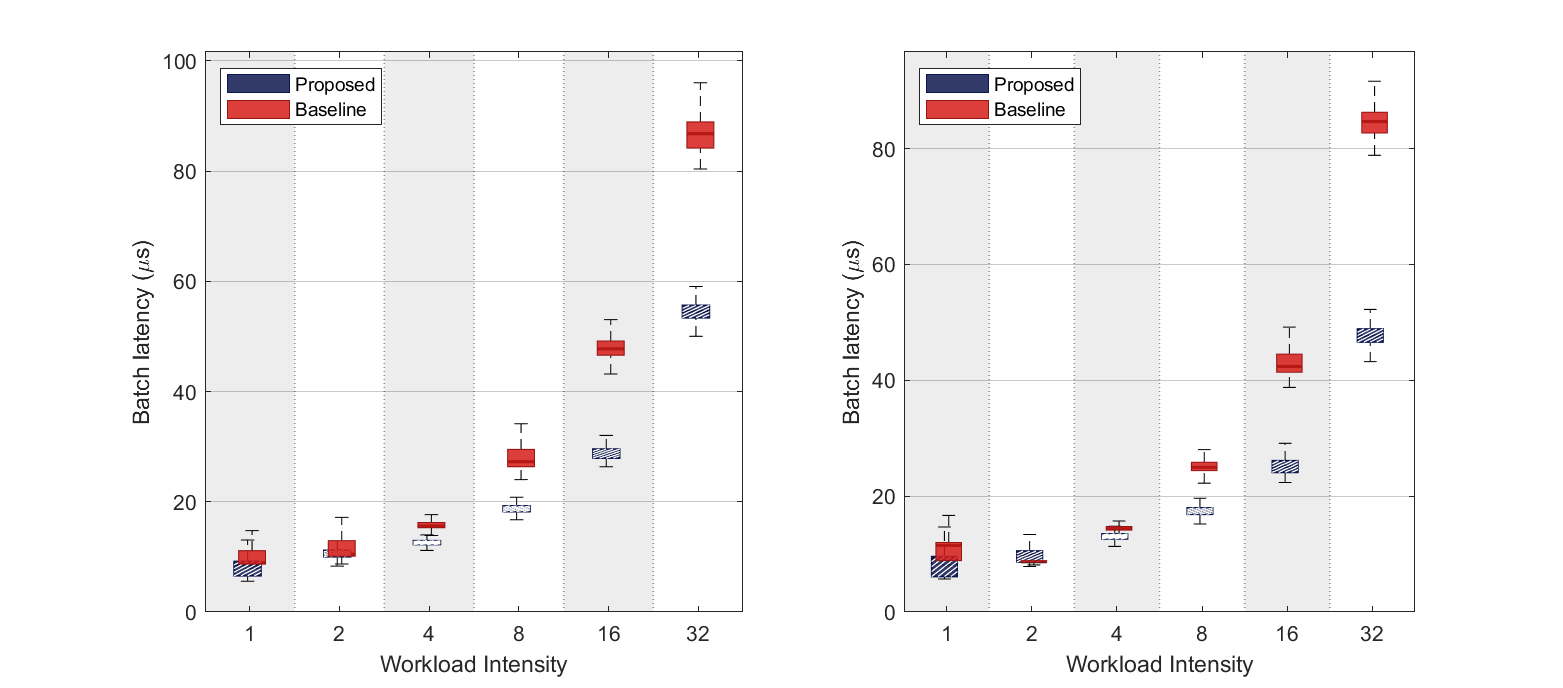}
\caption{Latency distribution versus workload intensity for (a) random read and (b) random write.}
\label{fig-box}
\end{figure}

\subsection{Latency Distribution}
Fig.~\ref{fig-box} shows the latency distribution as a function of workload intensity at a representative I/O size. Across the entire range, the proposed path exhibits consistently lower latency than the baseline, and the gap grows as the workload intensity increases. At an intensity of $32$, the median latency of the baseline reaches $87~\mu$s for random read and $85~\mu$s for random write, while the proposed path remains at $54~\mu$s and $48~\mu$s, respectively. The distributions are also tighter under the proposed path, indicating that coalescing per-device submissions reduces not only the mean latency but also its variability, consistently across reads and writes.

\subsection{Speedup Across Configurations}
\begin{figure}[t]
\centering
\includegraphics[width=3.8in, height=1.56in]{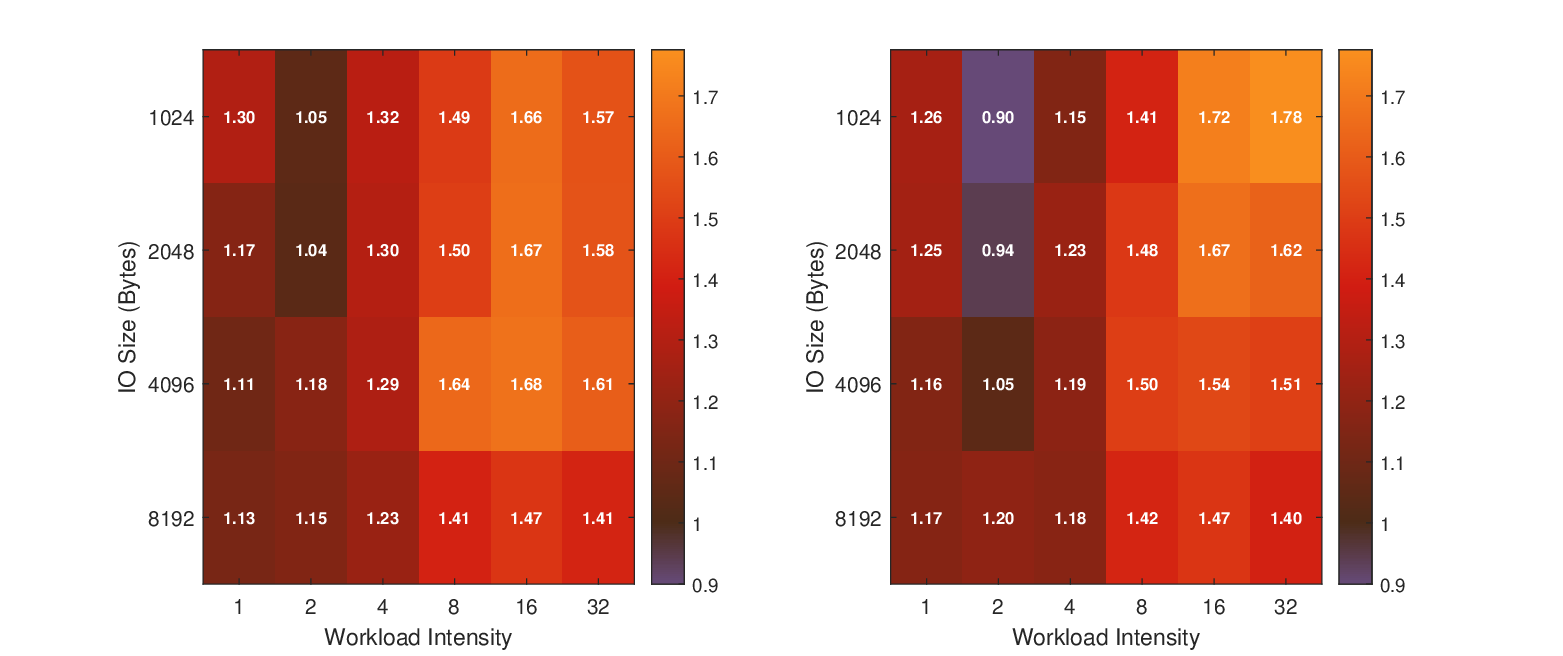}
\caption{Baseline-to-proposed mean latency ratio across I/O size and workload intensity for (a) random read and (b) random write.}
\label{fig-heat}
\end{figure}

Fig.~\ref{fig-heat} presents the baseline-to-proposed mean latency ratio over the I/O size and workload-intensity configurations. The ratio exceeds unity in nearly all configurations, confirming that the latency reduction generalizes across the configurations. Two effects drive the results. First, the ratio increases with workload intensity, reaching up to $1.78\times$, since the per-device round-trip overhead that the baseline incurs accumulates with the number of in-flight requests. Second, the ratio is larger for smaller I/O sizes, as the fixed per-request overhead dominates when the payload is small. For larger I/O, the ratio narrows toward $1.4\times$ as the path becomes bandwidth-bound. The ratio briefly decreases near an intensity of $2$, where too few requests are coalesced to offset the parsing and dispatch cost. These results indicate that the proposed mechanism is most effective for small-request, high-concurrency I/O, an access pattern common in data-intensive workloads scale with $N$.

\section{Conclusion}
We identified the per-device submission structure, rather than the kernel path alone, as a key factor in multi-device I/O overhead in virtualized data-center environments, and showed that it persists even in kernel-bypass stacks such as SPDK. To address this, we proposed a fan-out relocation mechanism in the SPDK vhost-user backend, in which the guest submits multi-device I/O as a single coalesced request that an acceleration bdev decomposes and dispatches per device. The evaluation indicated a latency improvement of up to $1.78\times$ over a per-device baseline, growing with the number of concurrent requests. These results suggest that backend-level coalescing is an effective direction for scaling storage I/O, and systems where one request fans out across multiple devices, relevant to data-intensive AI workloads. As future work, we will extend the mechanism to large-scale AI data centers with heterogeneous IP blocks.

\section*{Acknowledgments}
This work was supported by the Korea Institute of Energy Technology Evaluation and Planning(KETEP) and the Ministry of Climate, Energy \& Environment(MCEE) of the Republic of Korea (RS-2022-KP002860), and also supported by the National Research Foundation of Korea(NRF) grant funded by the Korea government(MSIT) (RS-2026-25485426).

\bibliographystyle{IEEEtran}
\bibliography{reference}
\newpage



\vfill

\end{document}